\documentstyle{article}
\newcommand{\beq}{\begin{equation}}
\newcommand{\eeq}{\end{equation}}
\newcommand{\beqa}{\begin{eqnarray}}
\newcommand{\eeqa}{\end{eqnarray}}
\newcommand{\bea}{\begin{eqnarray}}
\newcommand{\eea}{\end{eqnarray}}
\newcommand{\opt}{\sigma_{{\rm min}}}
\newcommand   {\etal}    {{\it et~al.}}
\input{psfig}
\begin{document}
\begin{titlepage}
\begin{flushright}
Revised version\\
LU TP 96-01\\
\today \\
\end{flushright}
\vspace{0.8in}
\LARGE
\begin{center}
{\bf Binary Assignments of Amino Acids\break from Pattern Conservation}\\
\vspace{.3in}
\large
Anders Irb\"ack\footnote{irback@thep.lu.se} and Frank Potthast\footnote{frank@thep.lu.se}\\
\vspace{0.05in}
Department of Theoretical Physics, University of Lund \\ 
S\"{o}lvegatan 14A,  S-223 62 Lund, Sweden \\
\vspace{0.15in}
To appear in {\sl Protein Engineering}
\end{center}
\normalsize
\vspace{1.2in}
\begin{center}
{\bf Abstract}
\end{center}
We develop a simple optimization procedure for assigning binary values to 
the amino acids. The binary values are determined by a maximization of the 
degree of pattern conservation in groups of closely related protein sequences.
The maximization is carried out at fixed composition. For compositions 
approximately corresponding to an equipartition of the residues, 
the optimal encoding is found to be strongly correlated with hydrophobicity. 
The stability of the procedure is demonstrated. 
Our calculations are based upon sequences in the SWISS-PROT database. 

\end{titlepage}
\newpage
\section{Introduction}

An amino acid sequence is a message in 20-letter code that determines the shape 
and function of the protein. This message is degenerate; amino acids may be 
exchanged to a certain degree without affecting the functionality of the 
protein in a drastic way~\cite{bowie}. For example, it has been demonstrated 
that the function of $\lambda$ repressor is very tolerant to exchanges 
of core residues as long as the pattern of hydrophobicity remains 
unchanged~\cite{lim}. The nature of the degeneracy can be probed 
by analyzing the Dayhoff mutation matrix~\cite{dayhoff} with 
respect to conservation of different physico-chemical properties. 
Results of such studies convincingly show that hydrophobicity plays
a central role in the formation of protein structure~\cite{taylor}.
  
As a first-order approach to the structure of proteins, it may therefore be 
tempting to take a simple two-letter code where the residues are classified 
as either hydrophobic or hydrophilic. A two-letter code can contain much 
structural information, as shown by studies of simplified models such as 
the lattice-based HP model~\cite{lau} or the off-lattice model of 
Refs.~\cite{stillinger,irback1}. However, there are $2^{20}$ possible ways 
to reduce a 20-letter code to a binary code, and it is not obvious that the  
most efficient encoding is closely related to a specific physico-chemical
property such as hydrophobicity.  

In this note we present a simple optimization procedure for assigning 
binary values, $\sigma_i=\pm1$, to the amino acids. The quantity which is
optimized is the degree of pattern conservation in groups of closely related 
proteins sequences. In order to avoid trivial solutions, the optimization is 
carried out for a fixed number of $\sigma_i=+1$, $N_+$. Depending on $N_+$, the 
optimal code may or may not have a simple physico-chemical interpretation.  
Here we shall focus on the results for $N_+=10$, which turn out to be 
strongly correlated with hydrophobicity. Although not unexpected
(cf Ref.~\cite{taylor}), this finding underlines the importance of 
hydrophobicity as the method is free from physico-chemical inputs. 
Furthermore, it should be stressed that the method is global in the sense 
that all possible encodings, for fixed $N_+$, are considered.      
  
Our calculations are based on groups of protein sequences extracted from 
the SWISS-PROT database \cite{swiss-prot}, each group corresponding to
a fixed length and a single protein but different species. Using a simple 
measure of similarity within these groups, the degree of pattern conservation
is maximized. Although the procedure ignores problems due to insertions 
and deletions of residues, it turns out to be fairly robust. The robustness 
was tested by separately analyzing different parts of the data set. 
In this way one can also study the stability of individual binary values, 
and assign refined, non-binary, values to the residues. 

Our method is somewhat related to the method of ``optimal matching 
hydrophobicities'' by Sweet and Eisenberg~\cite{sweet}. In this method   
hydrophobicity values are determined from mutation probabilities~\cite{dayhoff}
by using an iterative procedure. Our method uses mutation frequencies rather
than probabilities, and has the advantage that there is no need to specify 
initial values.

Another optimization procedure for determining hydrophobicity values  
has been proposed by Cornette et al.~\cite{cornette}. This method uses 
secondary-structure data rather than mutation data; sequence segments  
that form $\alpha$-helices are analyzed using Fourier methods. The  
hydrophobicity scale is obtained by maximizing of the strength of
the signal for the 3.6 residue period characteristic of the $\alpha$-helix. 
A similar method has been used for detecting patterns in biologically related 
sequences~\cite{viari}.

The paper is organized as follows. In Sec.~2 we describe our method.
In Sec.~3 we discuss the stability of the method and the results obtained.
We end with a brief summary in Sec.~4.
\section{Method}
\subsection{Forming groups of sequences}
\begin{figure}[tbp]
\begin{center}
\vspace{-35mm}
\mbox{\hspace{0mm}\psfig{figure=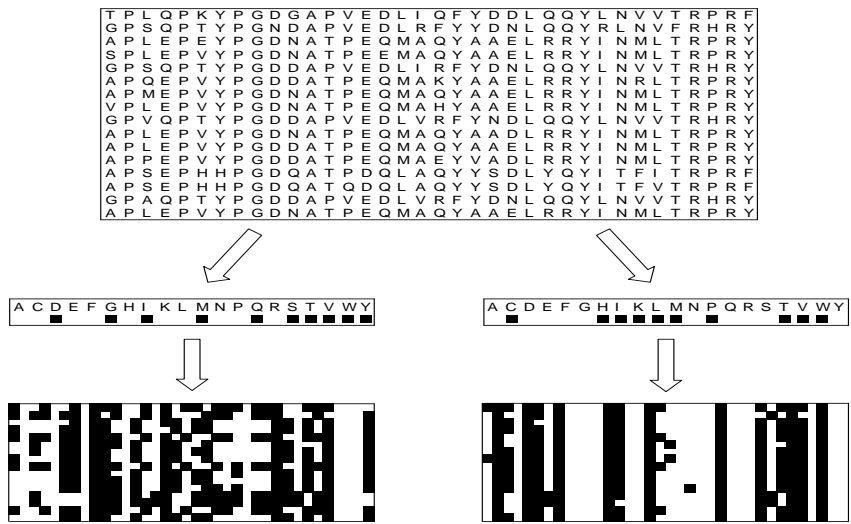,width=14.5cm,height=14cm}}
\vspace{-45mm}
\hspace{80mm}
\end{center}
\caption{Illustration of the method for one of the sequence groups.
The group consists of 16 sequences of pancreatic hormone with $N=36$,
which are shown in the upper box. Shown below are two binary
assignments and the resulting patterns.}
\label{fig:1}
\end{figure}

In our calculations we have used a set of sequence groups extracted from the 
SWISS-PROT database, release 31~\cite{swiss-prot}. Each group consisted  
of sequences with the same protein name and length $N$, but different
biological sources. All possible groups of this type were formed 
for $N\le 140$. As an example, one of the groups is shown in Fig.~\ref{fig:1}. 
In order to test the size dependence of the results, the data set was divided 
into two parts corresponding to $N\le 100$ (380 groups containing 1251 sequences 
in total) and $100<N\le 140$ (227 groups, 717 sequences), respectively. 

In forming these sequence groups we have ignored problems related to insertions and 
deletions of residues. This was done in order to keep the
method as free as possible from physico-chemical inputs. It is possible that
the method can be improved by incorporating some carefully chosen alignment 
technique. Also, the groups are small; typically, they contain three to
four sequences. It is therefore important to check the stability of the
procedure, which will be done in Sec.~3.

In our analysis we have removed uncertain sequences by ignoring
all entries in the database containing the feature keys
{\bf UNSURE} (indicates uncertainties in the sequence), {\bf NON\_TER} 
(an extremity of the sequence is not the terminal residue) and {\bf NON\_CONS}
(two residues in the sequence are not consecutive). Furthermore, we removed 
sequences containing the letters {\bf B} (denoting Asp or Asn),
{\bf Z} (Gln or Glu) and {\bf X} (any amino acid).
\subsection{Comparing binary assignments}
A binary encoding $\sigma$ of the amino acids is a mapping from the original 
20-letter alphabet to a two-letter alphabet, which we shall write as 
$\xi\to\sigma(\xi)\in \{+ 1 , -1\}$ where $\xi$ specifies the amino-acid type.  
Using the sequence groups described above, we shall assign a 
number $\Delta^\sigma$ to each encoding $\sigma$. The quantity $\Delta^\sigma$ 
provides a measure of pattern conservation and will be optimized. 
To define $\Delta^\sigma$ we proceed in three steps. 

First we define the distance $D^{\sigma}(\xi^a,\xi^b)$ between two 
arbitrary amino-acid sequences $\xi^a = ( \xi_{1}^a,\ldots,\xi_{N}^a)$ 
and $\xi^b = (\xi_{1}^b,\ldots,\xi_{N}^b)$ of the same length $N$,   
\begin{equation}
D^{\sigma}(\xi^a,\xi^b) = \sum_{i=1}^{N} 
1 - \delta_{\sigma (\xi^a_{i}),\sigma (\xi^b_{i})}
\qquad\qquad
\delta_{st}=\left\{\begin{array}{ll}
                   1 & \mbox{if $s=t$}\\
                   0 & \mbox{otherwise}
                   \end{array}
            \right.   
\end{equation}
$D^\sigma(\xi^a,\xi^b)$ is the usual Hamming distance between the binary strings 
$(\sigma(\xi_1^a),\ldots,\sigma(\xi_N^a))$ and\break
$(\sigma(\xi_1^b),\ldots,\sigma(\xi_N^b))$ and takes integer values between 
0 and $N$. Note that $D^\sigma(\xi^1,\xi^2)$ remains unchanged under a 
simultaneous change of the signs of all $\sigma(\xi)$'s, 
i.e., $D^{\sigma}(\xi^{1},\xi^{2})=D^{\sigma^{-}}(\xi^{1},\xi^{2})$
for all $\xi^1$ and $\xi^2$ if $\sigma^-(\xi)=-\sigma(\xi)$ for all $\xi$.
This implies a twofold degeneracy in $\sigma$ space. 

Next we define a measure of the flucuations within one group of sequences.
For a group labeled $k$ consisting of the $P_{k}$ 
sequences $\xi^{1},\ldots,\xi^{P_{k}}$ (all having the same length) we put
\begin{equation}
D^{\sigma}_k = \frac{1}{P_k}\sum_{1\le a<b\le P_k}D^{\sigma}(\xi^a,\xi^b)
\end{equation}
where the normalization is chosen so as to make the scaling of $D^{\sigma}_k$ 
linear in $P_{k}$ (the sum has $P_{k} (P_{k}-1) /2$ terms). A low 
$D^\sigma_k$ value signals high degree of similarity between the 
binary strings. The calculation of $D^\sigma_k$ is
illustrated in Fig.~\ref{fig:1} for two different $\sigma$; 
$D^\sigma_k$ is high to the left and low to the right.

The quantity $D^\sigma_k$ will not have a unique minimum $\sigma=\opt$ 
unless the group labeled $k$ is large. Finally, we therefore take a set 
of $f$ different sequence groups and define the optimal encoding as
\beq
\opt=\min_\sigma \Delta^\sigma
\eeq 
where 
\begin{equation}
\Delta^{\sigma} = \sum_{k=1}^{f}D^{\sigma}_k
\end{equation}

It is not meaningful to minimize $\Delta^\sigma (\ge 0)$ over all possible 
$\sigma$, since $\Delta^\sigma$ vanishes if $\sigma(\xi)$ is a constant. 
In order to avoid this trivial solution, we have performed minimizations 
for different fixed numbers of positive $\sigma(\xi)$'s, $N_+$. The  
twofold degeneracy mentioned above disappears when $N_+$ is held fixed and 
$N_+\ne10$. Furthermore, the symmetry implies that it is sufficient to 
consider $N_+ \leq 10$. The number of distinguishable encodings 
varies from 20 for $N_+=1$ to 92378 for $N_+=10$.
\section{RESULTS}
\subsection{Minimizing $\Delta^{\sigma}$}
We have performed the minimization described in the previous section 
for $N\le100$ and $100<N\le140$ and for $N_+=1,\ldots,10$. In this subsection
we present the results and discuss the robustness of the procedure. 
In the next subsection we discuss the interpretation of the results.

The minimizations were carried out in a straightforward way by enumerating
all possible encodings, which turned out to be feasible for all $N_+$.    
In Table~\ref{tab:1} we show the minimum of $\Delta^\sigma$, $\opt$, for 
different $N$ and $N_+$. The interpretation of $\opt$ depends on $N_+$,
as will be discussed below. Results for different $N_+$ are therefore 
not to be thought of as corresponding to some fixed property such 
as hydrophobicity (the
results for $N_+=10$ are strongly correlated with hydrophobicity, as 
will be shown below).

\begin{table}[tbp]
\begin{center}
\begin{tabular}{|c|cccccccccccccccccccc|}\hline
$N_+$   & A & C & D & E & F & G & H & I & K & L & M & N & P & Q & R & S & T & V & W & Y \\
 \hline 
   1    &$-$&$-$&$-$&$-$&$-$&$-$&$-$&$-$&$-$&$-$&$-$&$-$&$-$&$-$&$-$&$-$&$-$&$-$& + & $-$\\ 
   2    &$-$&$-$&$-$&$-$&$-$&$-$& + &$-$&$-$&$-$&$-$&$-$&$-$&$-$&$-$&$-$&$-$&$-$& + & $-$\\ 
   3    &$-$&$-$&$-$&$-$&$-$&$-$& + &$-$&$-$&$-$&$-$&$-$&$-$&$-$&$-$&$-$&$-$&$-$& + &  + \\ 
   4    &$-$&$-$&$-$&$-$&$-$&$-$& + &$-$&$-$&$-$& + &$-$&$-$&$-$&$-$&$-$&$-$&$-$& + &  + \\ 
   5    &$-$&$+$&$-$&$-$& + &$-$& + &$-$&$-$&$-$&$-$&$-$&$-$&$-$&$-$&$-$&$-$&$-$& + &  + \\ 
   6    &$-$& + &$-$&$-$& + &$-$& + &$-$&$-$&$-$& + &$-$&$-$&$-$&$-$&$-$&$-$&$-$& + &  + \\ 
   7    &$-$& + &$-$&$-$& + &$-$& + &$-$&$-$&$-$& + &$-$& + &$-$&$-$&$-$&$-$&$-$& + &  + \\ 
   8    &$-$&$-$&$-$&$-$& + &$-$& + & + &$-$& + & + &$-$&$-$&$-$&$-$&$-$&$-$& + & + &  + \\ 
   9    &$-$& + &$-$&$-$& + &$-$& + & + &$-$& + & + &$-$&$-$&$-$&$-$&$-$&$-$& + & + &  + \\ 
  10    &$-$& + &$-$&$-$& + &$-$& + & + &$-$& + & + &$-$& + &$-$&$-$&$-$&$-$& + & + &  + \\ 
 \hline 
   1    &$-$&$-$&$-$&$-$&$-$&$-$&$-$&$-$&$-$&$-$&$-$&$-$&$-$&$-$&$-$&$-$&$-$&$-$&$+$&$-$\\       
   2    &$-$& + &$-$&$-$&$-$&$-$&$-$&$-$&$-$&$-$&$-$&$-$&$-$&$-$&$-$&$-$&$-$&$-$& + &$-$\\       
   3    &$-$& + &$-$&$-$&$-$&$-$& + &$-$&$-$&$-$&$-$&$-$&$-$&$-$&$-$&$-$&$-$&$-$& + &$-$\\       
   4    &$-$& + &$-$&$-$&$-$&$-$& + &$-$&$-$&$-$&$-$&$-$&$-$&$-$&$-$&$-$&$-$&$-$& + & + \\       
   5    &$-$& + &$-$&$-$& + &$-$& + &$-$&$-$&$-$&$-$&$-$&$-$&$-$&$-$&$-$&$-$&$-$& + & + \\       
   6    &$-$& + &$-$&$-$& + &$-$& + &$-$&$-$&$-$& + &$-$&$-$&$-$&$-$&$-$&$-$&$-$& + & + \\       
   7    &$-$& + &$-$&$-$& + &$-$& + &$-$&$-$&$-$& + &$-$&$-$& + &$-$&$-$&$-$&$-$& + & + \\       
   8    &$-$& + &$-$&$-$& + &$-$&$-$& + &$-$& + & + &$-$&$-$&$-$&$-$&$-$&$-$& + & + & + \\       
   9    &$-$& + &$-$&$-$& + &$-$& + & + &$-$& + & + &$-$&$-$&$-$&$-$&$-$&$-$& + & + & + \\       
  10    &$-$& + &$-$&$-$& + &$-$& + & + &$-$& + & + &$-$& + &$-$&$-$&$-$&$-$& + & + & + \\ 
 \hline
\end{tabular}

\caption{$\opt$ for different values of $N_+$. The results were obtained by using groups 
of length $N\le 100$ (top) and $100 < N \leq 140$ (bottom). For $N_+=10$ there are two 
symmetry-related copies of $\opt$, and we have chosen the one with $\sigma(A)=-1$.}
\label{tab:1}
\end{center}
\end{table}

From Table~\ref{tab:1} it can be seen that the size dependence of the results 
is fairly weak, and that the structure of $\opt$ varies 
slowly with $N_+$. These observations indicate a certain degree of stability. 
In order to check the stability of the method in more detail, we divided 
the data set into blocks (see Table~\ref{tab:2}) and performed a separate  
optimization for each block. In Table~\ref{tab:3} we show 
the results of these calculations for $N_+=10$. For each block the 
irrelevant overall sign of $\opt$ (see Sec.~2) was chosen so as to 
minimize the Hamming distance to $\opt$ for the full data set. For the full 
data set the overall sign was fixed by setting $\opt(A)=-1$. The results
shown in Table~3 clearly demonstrate the stability of the method;
for example, it can be seen that twelve of the amino acids have been assigned 
the same binary value in twelve or more of the fourteen independent block 
calculations. Note that the stability of the assignment is amino-acid 
dependent. This will be used below to define a refined, non-binary, scale. 

\begin{table}[tbp]
\begin{center}
\begin{tabular}{|c|cccc|}\hline
Block & First entry & Last entry & No. of sequences & No. of groups \\ 
\hline
B1            &   ACBP\_BOVIN         &  CRBL\_VESXA        &  158     & 51       \\
B2            &   CSPA\_ECOLI         &  FER\_SYNY4         &  161     & 45       \\
B3            &   FER\_ARCLA          &  LYS3\_SHISO        &  149     & 52       \\
B4            &   MAST\_POLJA         &  NXS2\_NAJNI        &  158     & 45       \\
B5            &   NXSB\_LATCR         &  PYY\_RANRI         &  169     & 40       \\
B6            &   REV\_SIVA1          &  S10A\_HUMAN        &  155     & 47       \\
B7            &   S10B\_BOVIN         &  VA15\_VARV         &  152     & 53       \\
B8            &   VB09\_VACCC         &  YVDC\_VACCV        &  149     & 47       \\
\hline
C1          &   ACP\_BRACM    &   CYC\_KLULA                &  117     &  33  \\
C2          &   CYC\_BRAOL    &   H3\_VOLCA                 &  119     &  39  \\
C3          &  H3\_ACRFO      &   NU3C\_WHEAT               &  121     &  36  \\
C4          &  NU3M\_ASCSU    &   RL20\_ECOLI               &  124     &  33  \\
C5          &  RL20\_MAIZE    &   VAL3\_TYLCV               &  116     &  41  \\
C6          &  VDBP\_CAMVC    &   YV1\_TYLCH                &  120     &  45  \\
\hline      
\end{tabular}
\caption{The blocks used for $N\le 100$ (top) and $100<N\leq 140$ (bottom).}
\label{tab:2}
\end{center}
\end{table}

\begin{table}[tbp]
\begin{center}
\begin{tabular}{|r|cccccccccccccccccccc|}\hline
$\xi$  & A & C & D & E & F & G & H & I & K & L & M & N & P & Q & R & S & T & V & W & Y \\ 
\hline \hline
Full &$-$& + &$-$&$-$& + &$-$& + & + &$-$& + & + &$-$& + &$-$&$-$&$-$&$-$& + & + & + \\ 
\hline
 B1 & + &$-$&$-$&$-$& + & + &$-$& + &$-$& + & + &$-$& + &$-$&$-$& + & + & + &$-$&$-$\\
 B2 &$-$&$-$&$-$&$-$& + &$-$& + & + &$-$& + & + &$-$& + &$-$& + &$-$&$-$& + & + & + \\
 B3 & + &$-$&$-$&$-$& + &$-$&$-$& + &$-$& + & + &$-$& + &$-$&$-$&$-$& + & + & + & + \\
 B4 & + &$-$&$-$&$-$& + & + &$-$& + &$-$& + & + &$-$& + &$-$&$-$& + & + & + &$-$&$-$\\
 B5 &$-$& + &$-$&$-$& + &$-$& + & + &$-$& + & + &$-$& + &$-$&$-$&$-$&$-$& + & + & + \\
 B6 &$-$& + &$-$&$-$& + &$-$& + & + &$-$& + & + &$-$& + &$-$&$-$&$-$&$-$& + & + & + \\
 B7 &$-$& + &$-$&$-$& + &$-$& + & + &$-$& + & + &$-$& + &$-$&$-$&$-$&$-$& + & + & + \\ 
 B8 &$-$& + &$-$&$-$& + &$-$& + &$-$& + &$-$& + & + &$-$& + & + &$-$&$-$&$-$& + & + \\ 
\hline \hline          
Full &$-$& + &$-$&$-$& + &$-$& + & + &$-$& + & + &$-$& + &$-$&$-$&$-$&$-$& + & + & + \\ 
\hline
 C1 &$-$& + &$-$&$-$& + &$-$& + & + &$-$& + & + &$-$&$-$&$-$& + &$-$&$-$& + & + & +  \\ 
 C2 &$-$& + &$-$&$-$& + &$-$& + & + &$-$& + & + &$-$&$-$& + &$-$&$-$&$-$& + & + & +  \\
 C3 &$-$& + &$-$&$-$& + &$-$& + & + &$-$& + & + &$-$& + &$-$&$-$&$-$&$-$& + & + & +  \\ 
 C4 & + &$-$&$-$&$-$& + &$-$&$-$& + &$-$& + & + &$-$& + &$-$&$-$& + & + & + & + &$-$ \\
 C5 &$-$& + &$-$&$-$& + &$-$& + & + &$-$& + & + &$-$& + &$-$&$-$&$-$&$-$& + & + & +  \\
 C6 & + & + &$-$&$-$& + &$-$&$-$& + &$-$& + & + &$-$&$-$&$-$&$-$&$-$& + & + & + & +  \\ 
\hline
\end{tabular}
\caption{$\opt$ as obtained using the full data set and the different blocks 
in Table~2 ($N_+=10$). The results in the upper part are for $N\le 100$, 
whereas those in the lower part are for $100<N\le140$.}
\label{tab:3}
\end{center}
\end{table}

\begin{figure}[tbp]
\begin{center}
\vspace{-51mm}
\mbox{\hspace{0mm}\psfig{figure=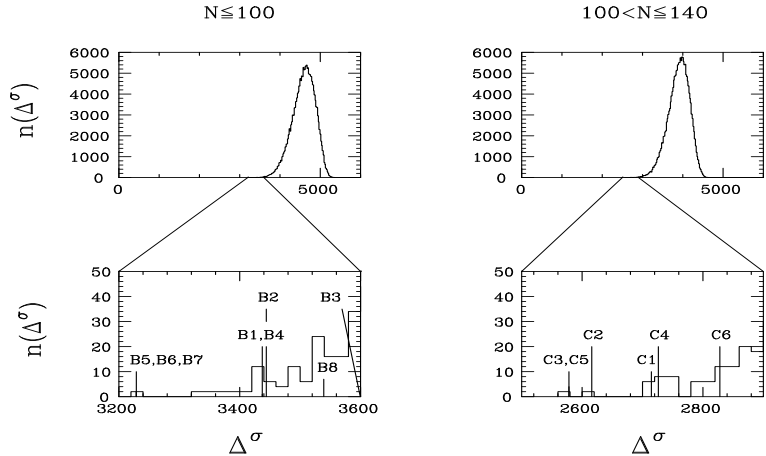,width=10.5cm,height=14cm}}
\vspace{-35mm}
\end{center}
\caption{a) Histograms of $\Delta^\sigma$ for $N_+=10$, obtained using
$N\leq 100$ (left) and $100 < N \leq 140$ (right). The lower figures 
are enlargements of the low-$\Delta^\sigma$ tails, in which we have
indicated the values of $\Delta^\sigma$ for the different minima $\opt$ 
from the block analysis. The minima $\opt$ for the blocks 
B5, B6 and B7 coincide with the minimum obtained using the full data set. 
The same is true for the blocks C3 and C5.} 
\label{fig:2}
\end{figure}

Another way to test the stability is to study the distribution of 
$\Delta^\sigma$. In Fig.~\ref{fig:2} we show two histograms of $\Delta^\sigma$ 
corresponding to $N\le100$ and $100<N\le140$, respectively, for $N_+=10$.
These histograms were obtained using the full data sets. Also using 
the full data set, we then computed $\Delta^\sigma$ for each $\opt$ 
from the block analysis. The positions of these fourteen values are shown   
in Fig.~\ref{fig:2}. They are all located in the low 0.1\% tails of the 
histograms, which gives another demonstration of the stability of the method.  
 
\subsection{Interpretation}
It is well-known that physico-chemical properties can be extracted from 
mutation probabilities, as, e.g., in the method of Sweet and 
Eisenberg~\cite{sweet}. Our approach is different since it is based on mutation 
frequencies rather than probabilities. Somewhat surprisingly, it 
turns out that physico-chemical information can be obtained in this way also.     
 
In our method the optimized quantity $\Delta^\sigma$ is defined as the total 
number of pattern-breaking bits, without reference to the frequency of 
occurrence of the amino acids. In this way emphasis is put on the overall 
degree of pattern conservation rather than the values assigned to individual  
amino acids. Whether the individual amino-acid values still contain useful 
information is a priori unclear. A necessary condition for that is, of 
course, that the corresponding degree of pattern conservation is 
relatively high. If, on the other hand, the mutations are more random, 
then the assigned values tend to reflect the frequency of 
occurrence of the amino acids; the procedure then tends 
to put rare amino acids in the smallest group 
(corresponding to $\sigma=+1$ if $N_+<10$).

\begin{figure}[tbp]
\begin{center}
\vspace{-35mm}
\mbox{\hspace{-31mm}\psfig{figure=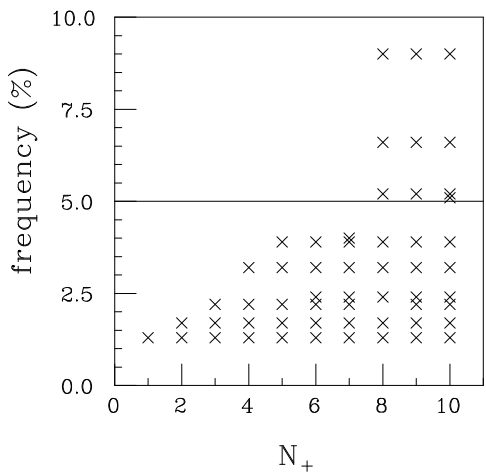,width=10.5cm,height=14cm}}
\vspace{-40mm}
\end{center}
\caption{Frequency of occurrence for the amino acids with $\sigma=+1$ 
plotted against $N_+$, using data for $100<N\le140$.} 
\label{fig:new}
\end{figure}

In Fig.~\ref{fig:new} we show amino-acid frequency~\cite{creighton} 
against $N_+$ for the results corresponding to $100<N\le140$. As can be
seen from this figure, there is a strong correlation between frequency and 
binary value for $N_+\le7$. This does not necessarily imply that the binary 
values solely reflect frequency; for example, W and C, the amino acids 
with $\sigma=+1$ for $N_+=2$ ($100<N\le 140$), are not only rare but have, in 
fact, been found to be the least mutable residues~\cite{jones}. However, it is 
clear that in order to extract any physico-chemical information for $N_+\le7$
a more detailed analysis is required, which is beyond the scope of the present 
paper.    

For $N_+\ge 8$ it is clear that the binary values contain non-trivial 
information since the correlation between binary value and frequency is
fairly weak. In what follows we shall discuss the results for $N_+=10$ 
in some detail.  

In order to take into account the fact that the stability of the binary 
value is amino-acid dependent, we begin by forming the average $a(\xi)$ of 
the results from the block analysis;  
\beq
a(\xi)={1\over k}\sum_{i=1}^k\opt^{(i)}(\xi)
\eeq
where $\opt^{(i)}$ denotes the result obtained using data block $i$.
As the size dependence was found to be weak, the average has been taken
over all the blocks in Table~\ref{tab:2} ($k=14$).  

\begin{figure}[tbp]
\begin{center}
\vspace{-25mm}
\mbox{\hspace{-31mm}\psfig{figure=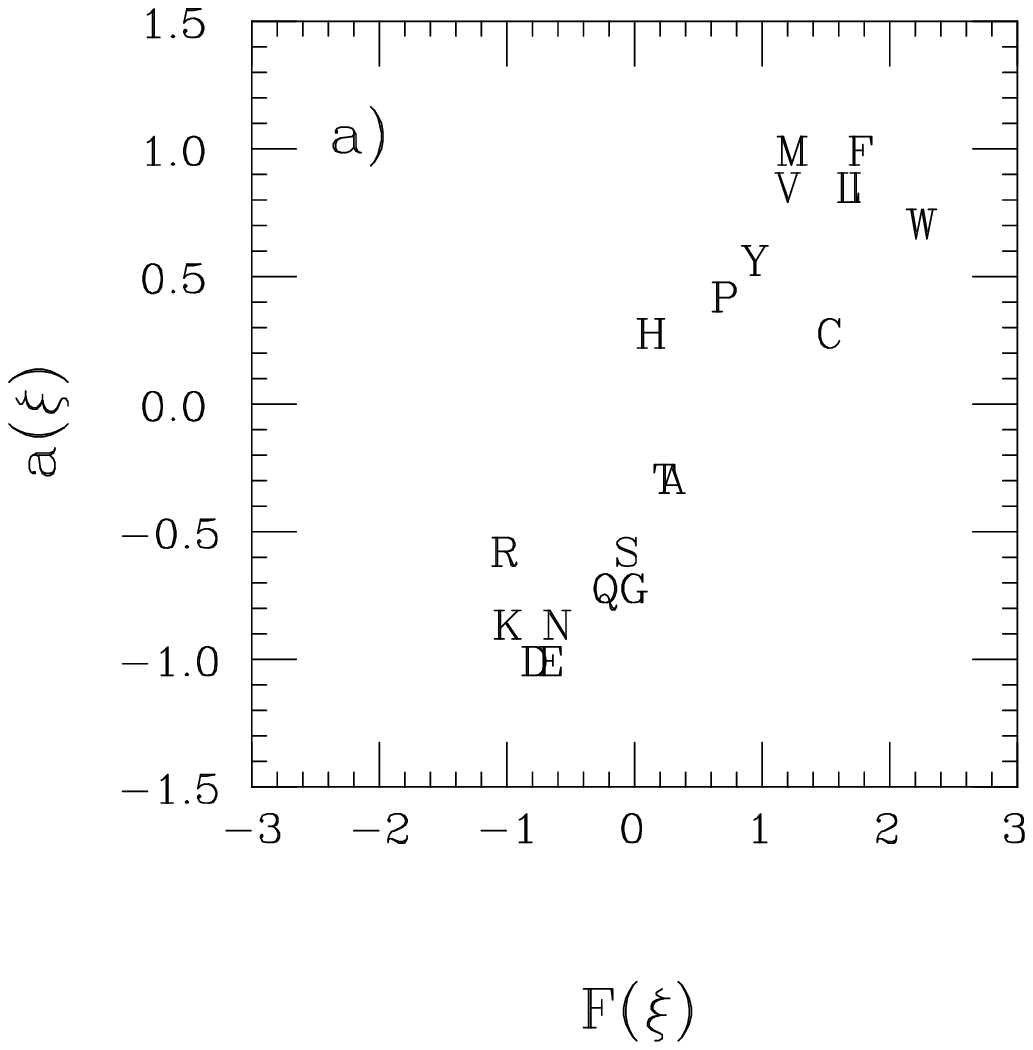,width=10.5cm,height=14cm}
\hspace{-30mm}\psfig{figure=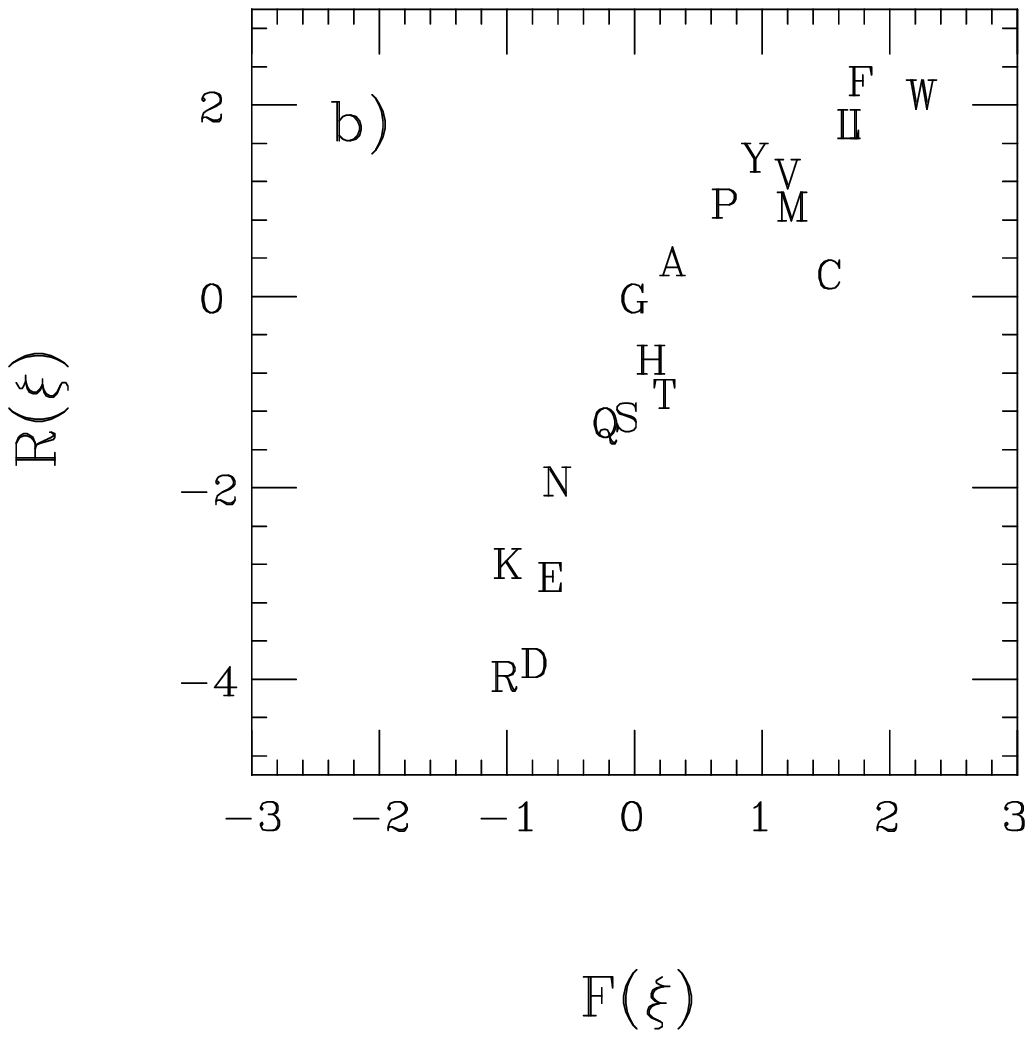,width=10.5cm,height=14cm}}
\vspace{-55mm}
\end{center}
\caption{a) $a(\xi)$ against the hydrophobicity scale of Fauch\`ere and 
Pliska~\protect\cite{fauchere}, $F(\xi)$. b) $F(\xi)$ against the 
hydrophobicity scale of Roseman~\protect\cite{roseman}, $R(\xi)$.}
\label{fig:3}
\end{figure}

In Fig.~\ref{fig:3}a we have plotted $a(\xi)$ against hydrophobicity,  
using the scale of Fauch\`ere and Pliska~\cite{fauchere}. 
From this figure it is clear that $a(\xi)$ is strongly correlated with
hydrophobicity. The correlation coefficient is 0.91, which, in fact, is
a representative value for what Cornette \etal\ typically found when comparing 
various existing hydrophobicity scales~\cite{cornette}. As an example, we 
show in Fig.~\ref{fig:3}b the scale of Fauch\`ere 
and Pliska against that of Roseman~\cite{roseman} (correlation coefficient
0.93). The correlation between $a(\xi)$ and hydrophobicity is much 
stronger than that between $a(\xi)$ and frequency of occurrence 
(correlation coefficient -0.32). 

The correlation between $a(\xi)$ and hydrophobicity becomes weaker as
$N_+$ decreases, but remains fairly strong down to $N_+=8$. 

%
\section{Summary}
We have developed a simple optimization procedure based on pattern conservation
for assigning binary values to the amino acids. In this method the optimal
encoding is determined by a global search at fixed composition, i.e., 
fixed value of the parameter $N_+$. The interpretation of the optimal 
encoding depends on $N_+$. For $N_+=10$ we have shown that the results are 
strongly correlated with hydrophobicity. Since the method is global and free 
from physico-chemical inputs, 
this finding illustrates the importance of the hydrophobicity pattern.

The stability of the method was demonstrated by applying it to independent 
sets of protein sequences. The method can probably be improved by 
incorporating some sequence alignment technique. However, it is important that 
this is done in such a way that the amount of physico-chemical input is kept 
at a minimum. The method can easily be generalized to non-binary scales.          

As an example of a possible application of our method, let us 
mention the question of how the statistical distribution of 
amino acids along functional protein sequences differs 
from a random distribution. This question has recently 
been addressed using binary assignments based on 
hydrophobicity~\cite{irback}. By studying long-range 
correlations, it was demonstrated that protein sequences differ from
random sequences in a statistically significant way. 
Alternative assignments such as those presented here may be useful
in studying the nature of these deviations from randomness.

{\bf Acknowledgements}\newline
We thank Carsten Peterson, Roberto Ugoccioni and Christian Witt for 
useful discussions.

\end{document}